%% file: letter1_v0.51.tex
\documentclass[final]{IEEEtran}
\pdfoutput=1
\usepackage{amsthm,amssymb,graphicx,multirow,amsmath,color,amsfonts}%,ulem}
\usepackage[update,prepend]{epstopdf}
\usepackage[noadjust]{cite}
\usepackage[latin1]{inputenc}
\usepackage{tikz}
\usetikzlibrary{arrows,calc}		% Optional ticks libraries
\usepackage{bbm} % for \mathbbm{1}
\usepackage{pdfpages}
\usepackage{tabulary}
\usepackage{multirow}
\usepackage{comment}
\usepackage{mathtools}
\usepackage{bigints}
\usepackage[inline]{enumitem}
\usepackage{adjustbox}
\include{notation}

\allowdisplaybreaks % Allows breaking of eqnarray over multiple pages (avoids unnecessary blanks in the document before eqnarray)

\begin{document}
\graphicspath{{./Figures/}}
\title{
Success Probability and Area Spectral Efficiency of a VANET Modeled as a Cox Process
}
\author{
Vishnu Vardhan Chetlur and Harpreet S. Dhillon
\thanks{The authors are with Wireless@VT, Department of ECE, Virginia Tech, Blacksburg, VA (email: \{vishnucr, hdhillon\}@vt.edu). The support of the US NSF (Grant IIS-1633363) is gratefully acknowledged.\hfill
Manuscript last updated: \today.
}
}
\maketitle

\begin{abstract}
This paper analyzes the performance of a vehicular ad hoc network (VANET) modeled as a Cox process, where the spatial layout of the roads is modeled by a Poisson line process (PLP) and the locations of nodes on each line are modeled as a 1D Poisson point process (PPP). For this setup, we characterize the success probability of a typical link and the area spectral efficiency (ASE) of the network assuming slotted ALOHA as the channel access scheme. We then concretely establish that the success probability of a typical link in a VANET modeled using a Cox process converges to that of a 1D and 2D PPP for some extreme values of the line and node densities. We also study the trends in success probability as a function of the system parameters and show that the optimum transmission probability that maximizes the ASE for this Cox process model differs significantly from those of the relatively-simpler 1D and 2D PPP models used commonly in the literature to model vehicular networks. % These insights offer useful design guidelines in the deployment of infrastructure to improve the performance of VANETs.
\end{abstract}
\begin{IEEEkeywords}
Stochastic geometry, Cox process, Poisson line process, Area spectral efficiency, Transmission probability.
\end{IEEEkeywords}

\section{Introduction} \label{sec:intro}

VANETs are one of the key building blocks of intelligent transportation systems. VANETs enable the vehicles and the road side units (RSUs) to share important local information such as the status of traffic, conditions of the roads and occurrence of accidents to apprise the drivers of potential hazards in advance~\cite{survey}. Since this information is time-sensitive, it imposes stringent connectivity and latency requirements on the network. Therefore, it is critical to understand the performance of VANETs to meet these design constraints. Unlike other popular wireless networks, VANETs have a peculiar spatial geometry as the locations of nodes are restricted to roadways. It is important to model this coupling between the location of nodes and the underlying road layout for an accurate assessment of the network performance. Given the irregularity in the node locations as well as the road layouts, one can treat this network as a stochastic network and use tools from stochastic geometry for the tractable performance analysis. Relevant prior art in this area is discussed next.

{\em Prior Art.}  Most works in the literature have adopted simpler spatial models such as 1D or 2D PPPs to model the locations of vehicular nodes and RSUs~\cite{5358011, busanelli}. In vehicular networks, the locations of nodes are restricted to roadways and hence, these models do not capture the strong coupling between the locations of nodes and the layout of roads. While there have been a few works in which this coupling was considered, the analysis was usually limited to a single road or a set of roads modeled by a deterministic set of lines~\cite{hesham, 8254665}. Although these works offer useful preliminary insights into the behavior of the network, they do not account for the irregularity in the spatial layout of roads. However, a few works have considered more sophisticated models such as a Cox process, where the road systems are modeled by a PLP~\cite{stoyan, baccplp} and further, the location of nodes are modeled by a 1D PPP~\cite{multihop, vishnuJ2, baccelliv2v}. In \cite{multihop}, the authors have studied the routing performance of a VANET for a linear multi-hop relay. In~\cite{vishnuJ2, baccelliv2v}, the authors have presented the canonical coverage analysis for nearest neighbor connectivity in this model. 
%The spatial coupling between the locations of nodes and the PLP makes the technical analysis quite challenging and hence, several performance metrics for a VANET, using this model, have not yet been studied in the literature. 
The main objective of this letter is to characterize the success probability of a typical link and spectral efficiency of a VANET modeled as a Cox process. More details of our contribution are provided next. 

{\em Contributions.} In this letter, we model the locations of nodes of a VANET by a Cox process where the road layout is modeled as a PLP and the locations of vehicular nodes and RSUs on each line (road) are modeled as independent homogeneous 1D PPPs. Assuming slotted ALOHA as the channel access scheme with a certain transmission probability for all the nodes in the network, we compute the success probability of a typical link and also the area spectral efficiency of the network. We then mathematically show that the success probability converges to that of 1D and 2D PPP models under specific (and practically relevant) asymptotic conditions. We also analyze the trends in success probability as a function of line and node densities and compare the optimum transmission probability that maximizes the ASE for a Cox process with those of well-investigated 1D and 2D PPPs. %These insights are quite useful in the deployment of infrastructure and design of protocols to improve the performance of VANETs.

\section{System Model}

\begin{figure}
	\centering
		\includegraphics[width=.23\textwidth]{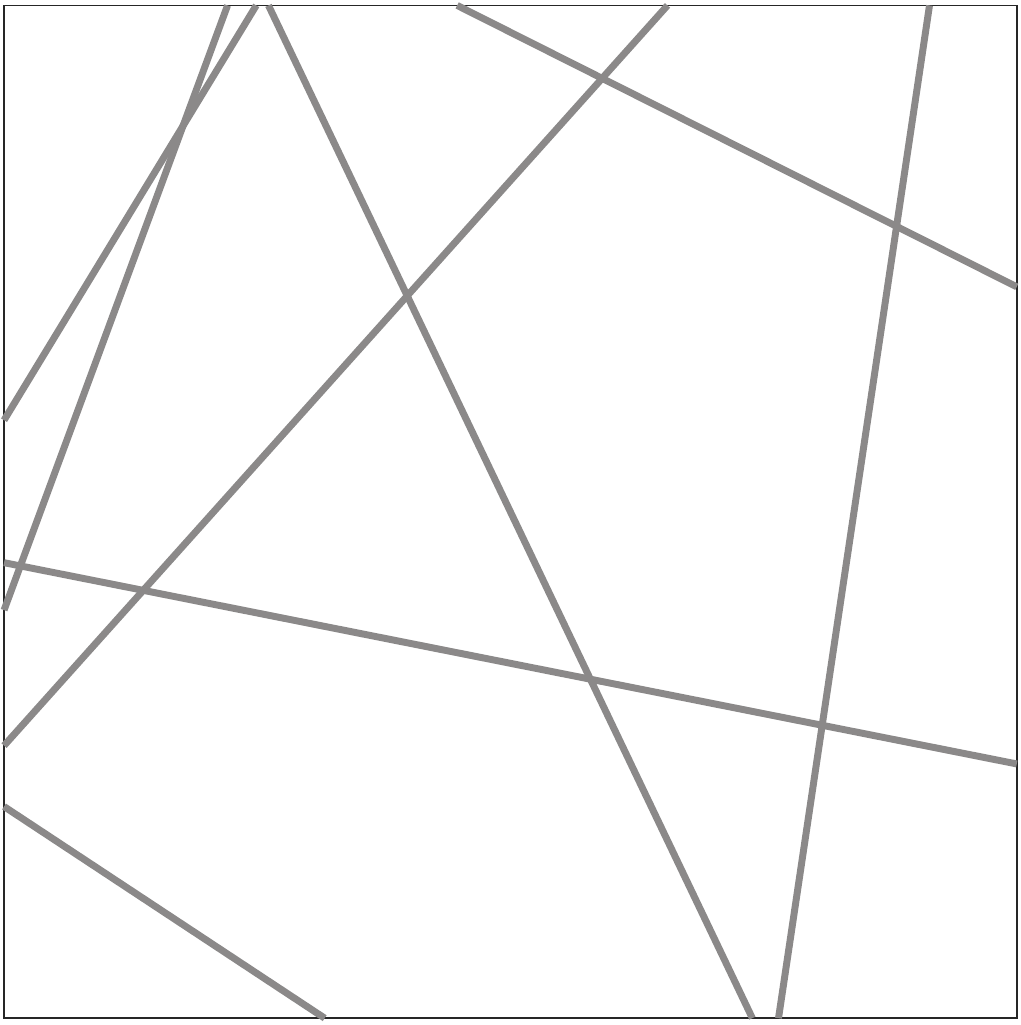}
		\includegraphics[width=.2315\textwidth]{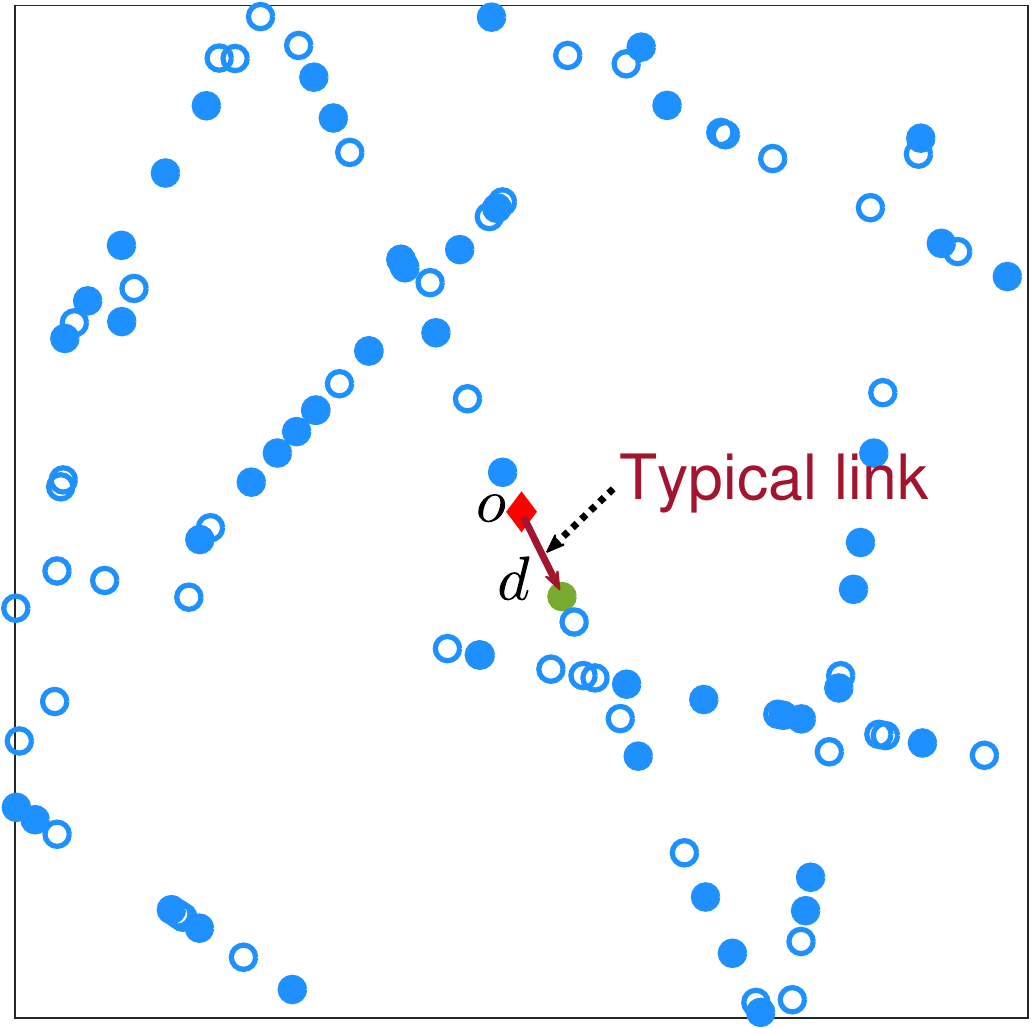}
	\caption{An illustration of PLP (left) and a Cox process driven by the PLP (right). The filled and hollow circles represent active and inactive transmitters in the network, respectively.}
	\label{fig:ptscox}
\end{figure}

We model the road systems which provide the physical support for the locations of vehicular nodes and RSUs by a PLP $\Phi_l$ with line density $\mu_l$. The density of the equivalent point process of $\Phi_l$ in the \textit{representation space} $\calC \equiv \nbbR^+ \times [0, 2\pi) $ is denoted by $\lambda_l \equiv \mu_l/ \pi$. We then model the location of vehicular nodes and RSUs on each line by homogeneous 1D PPPs with densities $\lambda_n$ and $\lambda_r$, respectively. Thus, the locations of the nodes on each line form a 1D PPP with density $\lambda_v = \lambda_n + \lambda_r$. We assume a slotted ALOHA channel access scheme where each node transmits independently of the other nodes in the network with a probability $p$. Therefore, the locations of transmitting nodes on each line $L$, denoted by $\Psi_L \equiv \{ {\rm w}_L\}$, at any instant form a thinned PPP with homogeneous density $p \lambda_v$. 
%Thus, the locations of transmitting nodes in the network form a Cox process $\Phi_t \equiv \{\Psi_{L_j}\}_{L_j \in \Phi_l}$ driven by the PLP $\Phi_l$ as shown in Fig. \ref{fig:ptscox}.
We further assume that the receiver nodes are located at a distance $d$ from their respective transmitting nodes on the same line. So, this forms a \textit{Cox bipolar network}, where the locations of the transmitting nodes form a Cox process $\Phi_t \equiv \{\Psi_{L_j}\}_{L_j \in \Phi_l}$ driven by the PLP $\Phi_l$ as illustrated in Fig. \ref{fig:ptscox}. It should be noted that Poisson bipolar networks (where transmitting nodes form a PPP) have been extensively used for the analysis of ad hoc networks in the past~\cite{haenggi2013stochastic}. Our goal is to characterize the signal-to-interference plus noise ratio (SINR) based success probability for a typical link of this Cox bipolar network. Without loss of generality, we assume that the receiver node of the typical link is located at the origin $o \equiv (0,0)$. {By applying Slivnyak's theorem~\cite{haenggi2013stochastic}, the translation of the origin can be interpreted as addition of a point at the origin to the PPP in the representation space $\calC$, thereby obtaining a PLP $\Phi_{l_0} \equiv \Phi_l \cup \{L_0\}$, where $L_0$ denotes the line containing the typical receiver located at the origin. Therefore, under Palm distribution, the resulting point process driven by the PLP $\Phi_{l_0}$ is the superposition of the Cox process $\Phi_t$ and a 1D PPP on the line $L_0$~\cite{morlot}. The line $L_0$ will henceforth be referred to as the typical line.} 
%Thus, the resulting point process, representing the locations of transmitting nodes, is the superposition of a 1D PPP on a line $L_0$ passing through the origin and the Cox process $\Phi_t$~\cite{morlot}. The line $L_0$ passing through the origin will henceforth be referred to as the typical line.

We assume that all the nodes transmit at the same power $P_t$. We further assume that all the communication links suffer from Rayleigh fading and the fading gains are exponentially distributed with mean 1. We consider a single-slope path-loss model with path-loss exponent $\alpha > 2$. Thus, the SINR at the receiver node of the typical link is given by
\begin{align}
\sinr =\frac{ P_t h_0 d^{-\alpha}}{\sum_{L_j \in \Phi_{l_0}} \sum_{{\rm w}_{L_j} \in \Psi_{L_j} } P_t  h_{{\rm w}_{L_j}} \|{\rm w}_{L_j}\|^{-\alpha} + \sigma^2},
\end{align}
where $h_0$ is the channel fading gain of the typical link, $h_{{\rm w}_{L_j}}$ is the channel fading gain between the receiver node of the typical link and the interfering node at the location ${\rm w}_{L_j}$, and $\sigma^2$ is the noise power.

\section{Success Probability}
This is the main technical section of the paper where we compute the SINR based success probability of the typical link. We will then analyze the asymptotic characteristics of the success probability for extreme values of line and node densities. We will also compute the ASE to characterize the overall performance of the network.

\subsection{Laplace Transform of Interference Distribution}
In this subsection, we derive the Laplace transform of the distribution of aggregate interference power at the receiver node of the typical link. We first categorize the sources of interference into two independent sets: (i) the set of nodes located on the typical line, and (ii) the set of nodes that are located on the lines other than the typical line. We denote the interference from these two components by $I_0$ and $I_1$, respectively. We will now characterize the interference from each of these components.

\begin{lemma}
	The Laplace transform of the distribution of interference power from the nodes located on the typical line is 
	\begin{align}\label{eq:lI0}
	\calL_{I_0} (s) =  \exp\bigg[-2 p \lambda_v \int_{0}^{\infty} \frac{1}{1 + (s P_t)^{-1} x^{\alpha}}  {\rm d}x \bigg].
\end{align}
\end{lemma}
\begin{IEEEproof}
The proof directly follows from the PGFL of 1D PPP and is hence skipped.
\end{IEEEproof}

\begin{lemma}
The Laplace transform of the distribution of interference power from all the nodes located on the lines other than the typical line is
\begin{align}\notag 
\calL_{I_1}(s)& = \exp \Bigg[  -2  \pi \lambda_l \int_0^{\infty} 1 - \exp \Big(	- 2 p \lambda_v\\ \label{eq:lI1}
& \hspace{2em} \times  \int_0^{\infty}  \frac{1}{1+ s^{-1} P_t^{-1} (y^2 + x^2)^{\alpha/2}} {\rm d} x \Big) {\rm d}y\Bigg].
\end{align}
\end{lemma}
\begin{IEEEproof}
	The Laplace transform of interference distribution can be computed as
\begin{align*}
&\calL_{I_1}(s) = \nbbE \bigg[ e^{-s \sum_{L_j \in \Phi_{l}} \sum_{{\rm w}_{L_j} \in \Psi_{L_j} } P_t h_{{\rm w}_{L_j}} \|{\rm w}_{L_j}\|^{-\alpha} }\bigg] \\
&\stackrel{(a)}{=}  \nbbE \bigg[ \prod_{L_j \in \Phi_{l}} \prod_{{\rm w}_{L_j} \in \Psi_{L_j} } \exp\big(-s P_t h_{{\rm w}_{L_j}} (y_j^2 + x_i^2)^{-\alpha/2} \big) \bigg] \\
%&= \nbbE_{\Phi_l} \bigg[ \prod_{L_j \in \Phi_{l}}  \nbbE_{\Psi_{L_j}} \bigg[ \prod_{{\rm w}_{L_j} \in \Psi_{L_j} } \frac{1}{1+ s (y_j^2 + x_i^2)^{-\alpha/2} } \bigg] \bigg]\\
&\stackrel{(b)}{=}   \nbbE_{\Phi_l} \bigg[ \prod_{L_j \in \Phi_{l}}  \exp \Big(	- 2 p \lambda_v \int_0^{\infty}  \frac{ P_t s}{P_t s+ (y_j^2 + x^2)^{\alpha/2}} {\rm d} x \Big) \bigg]\\
&\stackrel{(c)}{=} \exp \Bigg[  \scalebox{2.0}[1.0]{-}2 \pi \lambda_l \mspace{-6mu} \int\limits_0^{\infty} \! 1 \scalebox{2.0}[1.0]{-} \exp \bigg[ \scalebox{2.0}[1.0]{-}  \mspace{-6mu} \int\limits_0^{\infty} \mspace{-3mu} \frac{2 p \lambda_v P_t s}{P_t s+  (y^2 + x^2)^{\alpha/2}} {\rm d} x \bigg] {\rm d}y\Bigg],	
\end{align*}
where $y_j$ in (a) denotes the perpendicular distance of the line $L_j$ from the origin, $x_i$ is the Euclidean distance of a node on the line $L_j$ from the projection of the origin onto the line $L_j$, (b) follows from the PGFL of 1D PPP on each line $L_j$, and (c) follows from the PGFL formula for the 2D PPP representing the line process $\Phi_l$ in the representation space $\calC$~\cite{stoyan, vishnuJ2}.
\end{IEEEproof}
Since the two components of interference $I_0$ and $I_1$ are independent, the Laplace transform of the distribution of aggregate interference power is simply given by 
\begin{align}\label{eq:lI}
\calL_I(s) =  \calL_{I_0}(s) \calL_{I_1}(s).
\end{align}

\subsection{Success Probability}
The success probability is formally defined as the probability with which the SINR at the receiver node of the typical link exceeds a predetermined threshold $\beta$. Using the Laplace transform of the interference power distribution, the success probability is computed in the following theorem.
\begin{theorem}
	The success probability of the typical link $\pc$ is
\begin{align}\notag 
&\pc = \exp\bigg[ -\frac{\beta \sigma^2 d^{\alpha} }{P_t} -2 p\lambda_v \int_{0}^{\infty} \frac{\beta d^{\alpha} x^{-\alpha}}{1 + \beta d^{\alpha} x^{-\alpha}}  {\rm d}x  -2 \pi \lambda_l\\ \label{eq:pc}
&  \times  \int\limits_0^{\infty} \mspace{-6mu} 1 \scalebox{2}[1.0]{-}  \exp \bigg[	\scalebox{2}[1.0]{-}  2 p \lambda_v \int\limits_0^{\infty} \mspace{-6mu} \frac{ \beta d^{\alpha} (y^2 + x^2)^{-\alpha/2}}{1+ \beta d^{\alpha} (y^2 + x^2)^{-\alpha/2}} {\rm d} x \bigg] {\rm d}y\Bigg].
\end{align}
\end{theorem}
\begin{IEEEproof}
The success probability can be computed as 
\begin{align} \notag
\pc &= \P (\sinr > \beta) = \P \Big( \frac{ P_t h_0 d^{-\alpha}}{I + \sigma^2} > \beta \Big)\\
& =\exp\left({-\beta \sigma^2 d^{\alpha}/P_t}\right) \calL_{I}\left({\beta d^{\alpha}}/{P_t}\right).
\end{align}
Upon substituting \eqref{eq:lI0}, \eqref{eq:lI1}, and \eqref{eq:lI} in the above equation, we obtain the final expression for success probability.
\end{IEEEproof}

\subsection{Asymptotic Characteristics of Success Probability}
In this subsection, we analyze the success probability for some extreme values of the line and node densities. We show that the success probability of the typical link for this setup converges to that of 1D and 2D PPPs in the following Lemmas.

\begin{lemma}\label{lem:1dppp}
	As the density of lines in the network approaches zero ($\lambda_l \to 0$), the success probability of the typical link converges to that of a 1D PPP and is given by 
	\begin{align} \label{eq:pc1}
	\pc^{(1)} = \exp\bigg[\scalebox{2}[1]{-} \frac{\beta \sigma^2 d^{\alpha} }{P_t} \scalebox{2}[1]{-} 2 p \lambda_v \beta^{1/\alpha} d \frac{\pi}{\alpha} \csc\Big(\frac{\pi}{\alpha}\Big) \bigg].
	\end{align}
\end{lemma}
\begin{IEEEproof}
	{Following the same approach used in Lemma 17 of \cite{vishnuJ2}, this result can be easily obtained by} applying the limit $\lambda_l  \to  0$ on the expression for $\pc$ given in \eqref{eq:pc}.
	\end{IEEEproof}
%\begin{remark}
%	The expression given in \eqref{eq:pc1} in the previous lemma is nothing but the success probability of a 1D PPP~\cite{haenggi2013stochastic} with density $p \lambda_v$. Therefore, as the density of lines in the network approaches zero (sparse roads), the success probability of the typical link converges to that of a 1D PPP with the same node density.
%\end{remark}
%We will now derive the success probability for the setup for high line density and low node density.

\begin{lemma}\label{lem:2dppp}
	As the line density approaches infinity ($\lambda_l \to \infty$) and node density tends to zero ($\lambda_v \to 0$) while the overall density of nodes remains unchanged, the success probability of the typical link converges to that of a 2D PPP with the same node density and is given by
	\begin{align} \label{eq:pc2}
	&\pc^{(2)}\mspace{-3mu} = \mspace{-3mu}\exp\bigg[\scalebox{2}[1]{-} \frac{\beta \sigma^2 d^{\alpha} }{P_t} \scalebox{2}[1]{-}\pi^2 p \lambda_l \lambda_v \beta^{2/\alpha} d^2 \frac{2\pi}{\alpha} \csc\Big(\frac{2\pi}{\alpha}\Big)\bigg].
	\end{align}
%	where $\lambda$ is the overall density of nodes given by $\lambda = \pi p \lambda_l \lambda_v$
\end{lemma}
\begin{IEEEproof}
As the overall density of nodes in the network $\lambda = \pi p \lambda_l \lambda_v$ remains constant, the application of the two limits $\lambda_l \to \infty$ and $\lambda_v \to 0$ can be simplified to a single limit by substituting $\lambda_v = \lambda/(\pi p \lambda_l)$ in the expression in \eqref{eq:pc}.
Therefore, the success probability can now be computed as 
	\begin{align*}
	&\pc^{(2)}= \lim_{\lambda_l \to \infty} \pc \\ 
	&= \lim_{\lambda_l \to \infty} \exp\bigg[ \scalebox{2.0}[1.0]{-} \frac{\beta \sigma^2 d^{\alpha} }{P_t} \scalebox{2}[1.0]{-} \frac{2 \lambda f}{\pi \lambda_l}  \scalebox{2}[1.0]{-} 2 \pi \lambda_l   \int_0^{\infty }  \mspace{-6mu} 1 \scalebox{2}[1.0]{-} \exp \Big( \frac{- 2 \lambda g }{\pi \lambda_l}  \Big)  {\rm d}y\bigg],
	\end{align*}
	where
	\begin{align*}  
	&f = \int\limits_0^{\infty} \mspace{-6mu} \frac{\beta d^{\alpha} x^{-\alpha}}{1 + \beta d^{\alpha} x^{-\alpha}}  {\rm d}x   \text{ and } g = \int\limits_0^{\infty} \mspace{-6mu} \frac{ \beta d^{\alpha} (y^2 + x^2)^{-\alpha/2}}{1+ \beta d^{\alpha} (y^2 + x^2)^{-\alpha/2}} {\rm d} x .
	\end{align*}
	Applying the properties of limits, we get 
	\begin{align*}
	\pc^{(2)} &=  e^{\scalebox{2.0}[1.0]{-} \frac{\beta \sigma^2 d^{\alpha} }{P_t}} \exp \bigg[\lim_{\lambda_l \to \infty} -2 \pi \lambda_l \int_0^{\infty} \mspace{-9mu} 1 - \exp \Big( \frac{- 2 \lambda g }{\pi \lambda_l}  \Big)  {\rm d}y\bigg] \\
	&\stackrel{(a)}{=} e^{\scalebox{2.0}[1.0]{-} \frac{\beta \sigma^2 d^{\alpha} }{P_t}} \exp \bigg[  \lim_{\lambda_l \to \infty} \int_0^{\infty} \sum_{k=1}^{\infty} \frac{2 (-2 \lambda g)^k}{(\pi \lambda_l)^{k-1} k!} {\rm d}y\bigg]\\
	&\stackrel{(b)}{=} e^{\scalebox{2.0}[1.0]{-} \frac{\beta \sigma^2 d^{\alpha} }{P_t}}\exp \bigg[  \sum_{k=1}^{\infty} \int_0^{\infty}  \lim_{\lambda_l \to \infty} \frac{2 (-2 \lambda g)^k}{(\pi \lambda_l)^{k-1} k!} {\rm d}y\bigg]\\
	%&\stackrel{(c)}{=} e^{\scalebox{2.0}[1.0]{-} \frac{\beta \sigma^2 d^{\alpha} }{P_t}}  \exp \bigg[  \int_0^{\infty} -4 \lambda  g {\rm d}y \bigg]\\%+ \sum_{k=2}^{\infty} \int_0^{\infty} \lim_{\lambda_l \to \infty} \frac{ 2(-2 \lambda g)^k}{(\pi \lambda_l)^{k-1} k!} {\rm d}y\bigg] \\
	%&= e^{\scalebox{2.0}[1.0]{-} \frac{\beta \sigma^2 d^{\alpha} }{P_t}} \exp\bigg[ -4 \pi \p \lambda_l \lambda_v \\  
	%&\hspace{4em} \times \int_0^{\infty} \int_0^{\infty} \frac{ \beta d^{\alpha} (y^2 + x^2)^{-\alpha/2}}{1+ \beta d^{\alpha} (y^2 + x^2)^{-\alpha/2}} {\rm d} x {\rm d}y \bigg] \\
	&\stackrel{(c)}{=} \exp\bigg[ - \frac{\beta \sigma^2 d^{\alpha} }{P_t} -2 \pi^2 p \lambda_l \lambda_v  \int_0^{\infty}  \frac{ \beta d^{\alpha} r^{1-\alpha}}{1+ \beta d^{\alpha} r^{-\alpha}} {\rm d} r \Bigg],
		\end{align*}
where (a) follows from the Taylor series expansion of exponential function, (b) follows from switching the order of integral and summation operations and applying Dominated Convergence Theorem (DCT), and (c) follows from the limit of the integrand which evaluates to $0$ for all $k>1$, and then substituting $x\mspace{-5mu} =\mspace{-5mu} r \cos\theta$ and $y\mspace{-5mu} =\mspace{-5	mu} r \sin\theta$ in the resulting integral.% and simplification of the resulting integral.
\end{IEEEproof}
{The asymptotic results presented in Lemmas \ref{lem:1dppp} and \ref{lem:2dppp} correspond to practical scenarios of sparse layout of roads and sparse vehicular traffic in dense layout of roads, respectively.}

\subsection{Area Spectral Efficiency}
The area spectral efficiency is defined as the average number of bits successfully transmitted per unit time per unit bandwidth per unit area in the network. Assuming that all the transmitted symbols are from Gaussian codebooks, the ASE can be computed using Shannon's capacity formula as 
\begin{align}
\ase = \lambda \pc \log_2(1+\beta) \text{    bits/s/Hz/km$^2$},
\end{align}
where $\lambda$ is the average number of active transmitting nodes per unit area given by $\lambda = \pi \lambda_l p \lambda_v$.

 Recall that the transmission probability $p$ essentially determines the density of active nodes in the network. As the density of active nodes increases, the interference experienced at the receiver node of a typical link increases, thereby degrading the success probability, which in turn worsens the ASE. However, an increase in the node density linearly scales the number of concurrently active links in the network, thereby improving the ASE. Owing to this conflicting impact of transmission probability on the overall spectral efficiency, there exists an optimum value of transmission probability $p^*$ that maximizes the ASE of the network. Due to the complexity of the final expression for the success probability, it is not possible to obtain a closed-form expression for $p^*$ for the Cox bipolar model. That said, since the success probability asymptotically converges to that of 1D and 2D PPPs under specific conditions as shown in Lemmas \ref{lem:1dppp} and \ref{lem:2dppp}, respectively, the optimum transmission probabilities in those cases can be computed in closed form~\cite{haenggi2013stochastic}. Their expressions are given below:
\begin{align}
&\text{1D PPP:} \hspace{.5em} p^* = \frac{\alpha (2  \lambda_v d \beta^{1/\alpha})^{-1} }{\pi \csc(\pi/\alpha) },  \hspace{1em} \lambda_l \to 0,\\
&\text{2D PPP:} \hspace{.5em} p^* = \frac{ \alpha (\pi^2  \lambda_l \lambda_v d^2 \beta^{2/\alpha})^{-1}}{ 2 \pi \csc(2 \pi/\alpha)}, \hspace{1em} {\lambda_l \to \infty}, { \lambda_v \to 0} .
\end{align}

%
%\begin{remark}
%	The expression for success probability obtained in \eqref{eq:pc2} in the previous lemma is nothing but the success probability of a typical receiver in a 2D PPP~\cite{haenggi2013stochastic}. Therefore, as the line density $\lambda_l$ approaches infinity (dense roads) and node density $\lambda_v$ goes to zero (sparse nodes) such that the overall node density (average number of nodes per unit area) remains constant, the success probability of the typical receiver converges to that of 2D PPP with density $\lambda = \pi p \lambda_l \lambda_v$.
%\end{remark}

\section{Results and Discussion}
In this section, we verify the accuracy of our analytical results presented in the previous section by comparing them with the results from Monte-Carlo simulations. We will also analyze the impact of line and node densities on the success probability and ASE. %This analysis offers useful design insights in the deployment of infrastructure in order to achieve the desired network performance.
\subsection{Success Probability}
\begin{figure}\centering
	\includegraphics[width=.26\textwidth]{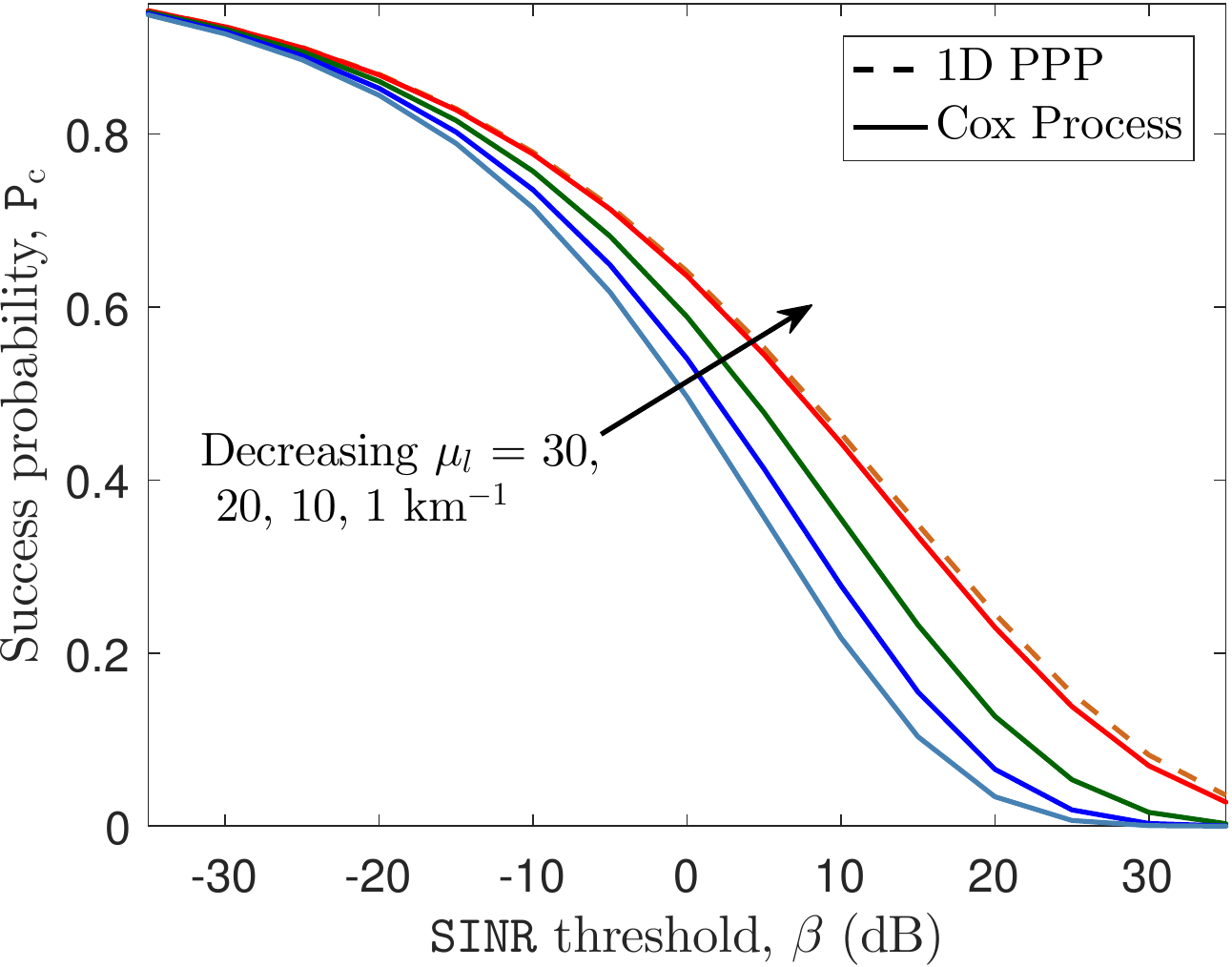}
	\caption{Success probability of the typical link as a function of SINR threshold ($\lambda_v = 20$ nodes/km, $p = 1$, $d = 0.01$ km, and $\alpha = 4$).}
	\label{fig:cov_ll}
\end{figure}
{\em Impact of line density}. We plot the success probability of the typical link as a function of $\sinr$ threshold for different line densities as shown in Fig. \ref{fig:cov_ll}. As the density of lines in the network increases, it reduces the distance between the receiver and the interfering nodes, thereby worsening the success probability. It can also be observed that the success probability corresponding to the Cox process aligns with that of a 1D PPP for a very low line density. %From Figs. \ref{fig:cov_lv} and \ref{fig:cov_ll}, we can say that the density of nodes has a stronger impact on the success probability than the density of lines in the network.

\begin{figure}\centering
	\includegraphics[width=.26\textwidth]{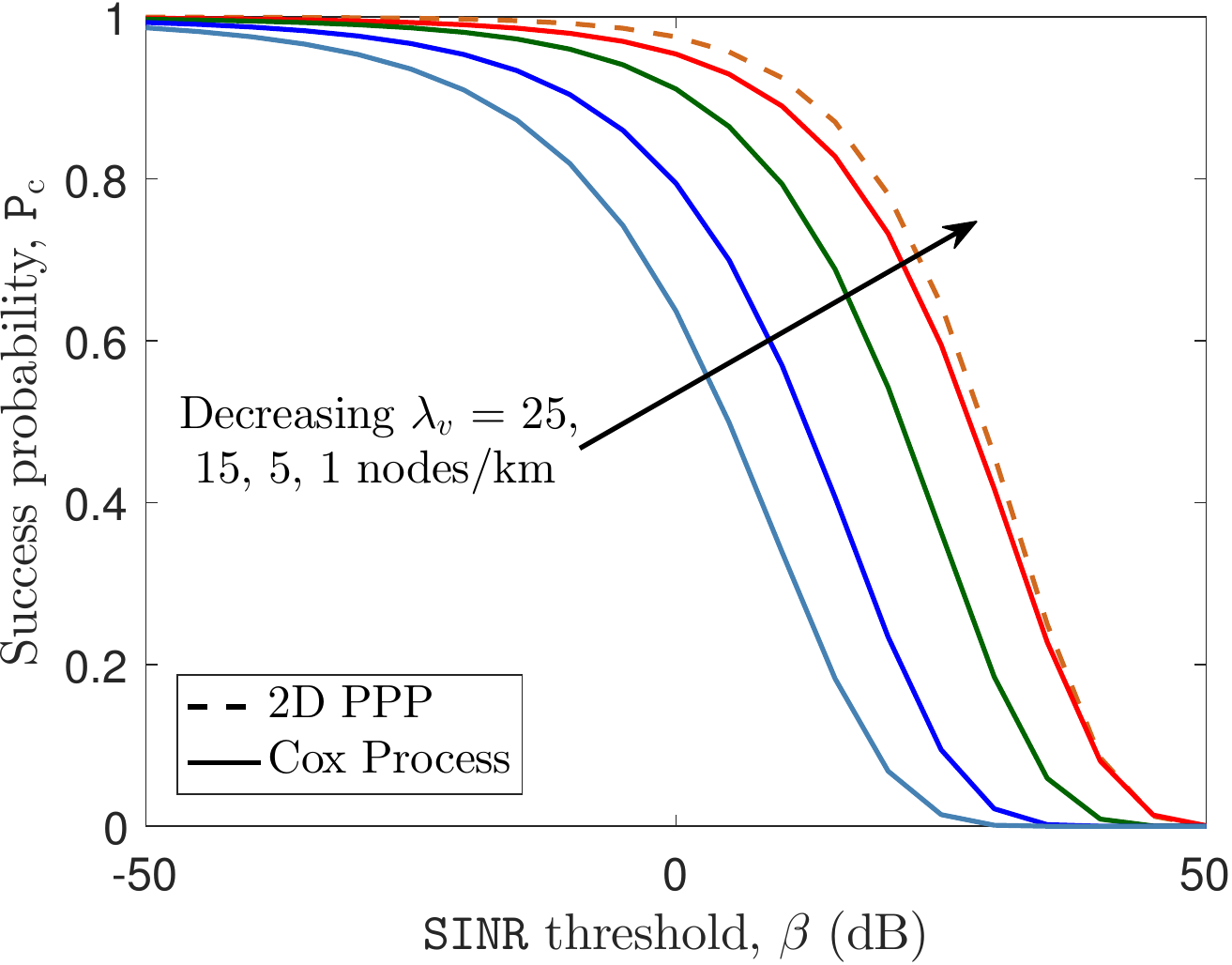}
	\caption{Success probability of the typical link as a function of SINR threshold for high line density ($\mu_l = 50$ km$^{-1}$, $d = 0.01$ km, $p=1$, and $\alpha = 4$).}
	\label{fig:cov_lv}
\end{figure}

{\em Impact of node density}.
We compare the success probability of the typical link for different values of node densities for a relatively high line density ($\mu_l = 50$ km$^{-1}$) and fixed value of transmission probability. As expected, the success probability worsens as the density of nodes increases due to the reduced distance between the receiver node and the interfering nodes, as shown in Fig. \ref{fig:cov_lv}. We also observe that the success probability corresponding to the Cox process aligns with that of a 2D PPP for low density of nodes on each line.
	\subsection{Optimum transmission probability to maximize ASE}
\begin{figure}[t]\centering
	\includegraphics[width=.26\textwidth]{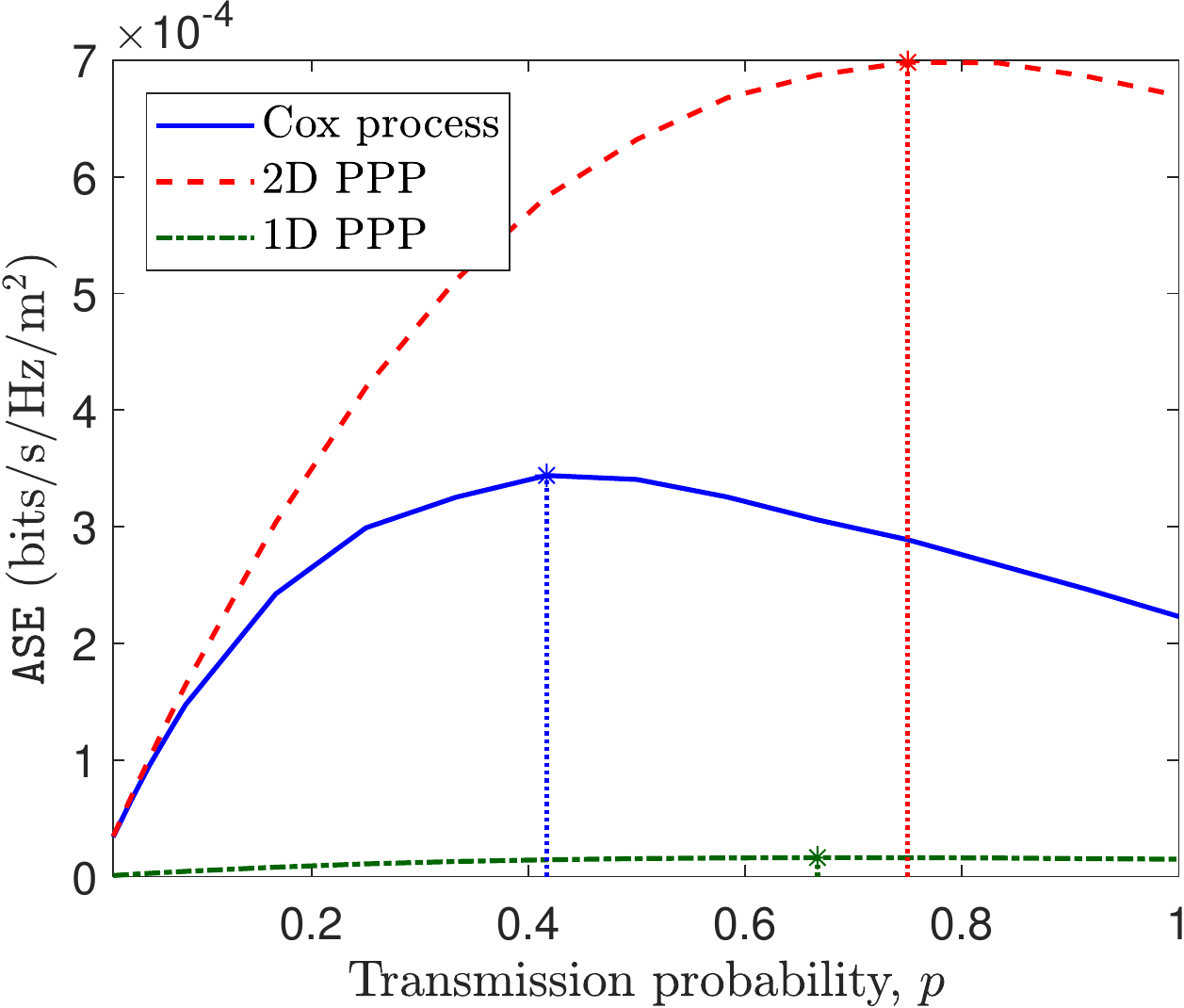}
	\caption{ASE of the network as a function of transmission probability $p$ ($\mu_l = 30$ km$^{-1}$, $\lambda_v = 60$ nodes/km, and $\alpha = 4$).}
	\label{fig:ase}
\end{figure}
We plot the ASE for the Cox bipolar model as a function of the transmission probability and compare it with that of canonical 1D and 2D PPP models as depicted in Fig. \ref{fig:ase}. While the locations of nodes are uniformly distributed in $\mathbb{R}^2$ in a PPP, they are confined to lie on the lines because of which the resulting point process exhibits ``clustering behavior''. Consequently, as the node density increases, the interference at the typical receiver increases at a faster rate in a Cox bipolar model than that of the Poisson bipolar model. Therefore, the optimum transmission probability that maximizes the ASE in a Cox bipolar model is smaller than that of the 1D and 2D PPP models with the same node density, as shown in Fig. \ref{fig:ase}.
% We know that the transmission probability $p$ essentially determines the density of active nodes in the network. We have already seen that the success probability of a link degrades as the density of nodes increases, which in turn worsens the spectral efficiency. However, an increase in the node density linearly scales the number of concurrently active links in the network, thereby improving the ASE. Owing to this conflicting impact of transmission probability on the overall spectral efficiency, there exists an optimum value of transmission probability $p^*$ that maximizes the ASE of the network as shown in Fig. \ref{fig:ase}. Since the success probability of the typical link asymptotically converges to that of 1D and 2D PPPs under certain conditions, the optimum transmission probabilities for these cases are given by
%\begin{align}
%&\text{1D PPP:} \hspace{.5em} p^* = \frac{(2 \alpha \lambda_v d \beta^{1/\alpha})^{-1} }{\pi \csc(\pi/\alpha) },  \hspace{1em} \lambda_l \to 0,\\
%&\text{2D PPP:} \hspace{.5em} p^* = \frac{ \alpha (\pi^2  p\lambda_l \lambda_v d^2 \beta^{2/\alpha})^{-1}}{ 2 \pi \csc(2 \pi/\alpha)}, \hspace{1em} {\lambda_l \to \infty}, { \lambda_v \to 0} .
%\end{align}
%However, for moderate values of line and node densities, the optimum transition probability and the corresponding ASE value for the Cox process differ significantly from those of the well-studied 1D and 2D PPP models as depicted in Fig. \ref{fig:ase}.

\section{Conclusion}
In this letter, we have analyzed the quality of a typical communication link in a VANET modeled as a Cox bipolar network. We have derived the exact expression for the success probability of the typical link assuming independent Rayleigh fading for all the links and also computed the area spectral efficiency of the network. We have mathematically shown that the success probability of the typical link asymptotically converges to that of 1D and 2D PPPs under certain conditions. Our analytical results yield an optimum transmission probability that maximizes the ASE of the network. These results could offer some useful insights into the deployment of RSUs to improve the performance of VANETs. A meaningful extension of this work could be to analyze the effects of shadowing on the performance of the network.

\bibliographystyle{IEEEtran}

\bibliography{letter1_v0.51.bbl}

\end{document}

%% file: notation.tex
\def\nb0{{\mathbf{0}}}
\def\nb1{{\mathbf{1}}}

\def\nbbE{{\mathbb{E}}}

\def\nbbR{{\mathbb{R}}}

\newtheorem{lemma}{Lemma}

\newtheorem{theorem}{Theorem}

\def\P{\mathbb{P}}
\def\pc{\mathtt{P_c}}
\def\sinr{\mathtt{SINR}}			% Signal to interference plus noise ratio

\def\ase{\mathtt{ASE}}

\def\calC{\mathcal{C}}

\def\calL{\mathcal{L}}

%% file: letter1_v0.51.bbl
\begin{thebibliography}{10}
\providecommand{\url}[1]{#1}
\csname url@samestyle\endcsname
\providecommand{\newblock}{\relax}
\providecommand{\bibinfo}[2]{#2}
\providecommand{\BIBentrySTDinterwordspacing}{\spaceskip=0pt\relax}
\providecommand{\BIBentryALTinterwordstretchfactor}{4}
\providecommand{\BIBentryALTinterwordspacing}{\spaceskip=\fontdimen2\font plus
\BIBentryALTinterwordstretchfactor\fontdimen3\font minus
  \fontdimen4\font\relax}
\providecommand{\BIBforeignlanguage}[2]{{%
\expandafter\ifx\csname l@#1\endcsname\relax
\typeout{** WARNING: IEEEtran.bst: No hyphenation pattern has been}%
\typeout{** loaded for the language `#1'. Using the pattern for}%
\typeout{** the default language instead.}%
\else
\language=\csname l@#1\endcsname
\fi
#2}}
\providecommand{\BIBdecl}{\relax}
\BIBdecl

\bibitem{survey}
H.~Hartenstein and L.~P. Laberteaux, ``A tutorial survey on vehicular ad hoc
  networks,'' \emph{IEEE Commun. Magazine}, vol.~46, no.~6, pp. 164--171, Jun.
  2008.

\bibitem{5358011}
B.~Blaszczyszyn, P.~{M\"{u}hlethaler}, and Y.~Toor, ``Maximizing throughput of
  linear vehicular {Ad-hoc NETworks} ({VANETs}) -- a stochastic approach,'' in
  \emph{European Wireless Conf.}, May 2009, pp. 32--36.

\bibitem{busanelli}
S.~Busanelli, G.~Ferrari, and R.~Gruppini, ``Performance analysis of broadcast
  protocols in {VANETs} with {Poisson} vehicle distribution,'' in \emph{Intl.
  Conf. on {ITS} Telecommunications}, Aug. 2011, pp. 133--138.

\bibitem{hesham}
M.~J. Farooq, H.~ElSawy, and M.~S. Alouini, ``A stochastic geometry model for
  multi-hop highway vehicular communication,'' \emph{IEEE Trans. on Wireless
  Commun.}, vol.~15, no.~3, pp. 2276--2291, Mar. 2016.

\bibitem{8254665}
J.~P. Jeyaraj and M.~Haenggi, ``Reliability analysis of {V2V} communications on
  orthogonal street systems,'' in \emph{Proc., IEEE Globecom}, Dec. 2017, pp.
  1--6.

\bibitem{stoyan}
S.~N. Chiu, D.~Stoyan, W.~S. Kendall, and J.~Mecke, \emph{Stochastic geometry
  and its applications}.\hskip 1em plus 0.5em minus 0.4em\relax John Wiley \&
  Sons, 2013.

\bibitem{baccplp}
F.~Baccelli, M.~Klein, M.~Lebourges, and S.~Zuyev, ``Stochastic geometry and
  architecture of communication networks,'' \emph{Telecommunication Systems},
  vol.~7, no.~1, pp. 209--227, Jun. 1997.

\bibitem{multihop}
B.~Blaszczyszyn and P.~Muhlethaler, ``Random linear multihop relaying in a
  general field of interferers using spatial {Aloha},'' \emph{IEEE Trans. on
  Wireless Commun.}, vol.~14, no.~7, pp. 3700--3714, Jul. 2015.

\bibitem{vishnuJ2}
V.~V. Chetlur and H.~S. Dhillon, ``Coverage analysis of a vehicular network
  modeled as {Cox} process driven by {Poisson} line process,'' {\em IEEE Trans.
  on Wireless Commun.}, to appear.

\bibitem{baccelliv2v}
C.~Choi and F.~Baccelli, ``An analytical framework for coverage in cellular
  networks leveraging vehicles,'' 2017, available online:
  arxiv.org/abs/1711.09453.

\bibitem{haenggi2013stochastic}
M.~Haenggi, \emph{Stochastic Geometry for Wireless Networks}.\hskip 1em plus
  0.5em minus 0.4em\relax Cambridge University Press, 2013.

\bibitem{morlot}
F.~Morlot, ``A population model based on a {Poisson} line tessellation,'' in
  \emph{Proc., Modeling and Optimization in Mobile, Ad Hoc and Wireless
  Networks}, May 2012, pp. 337--342.

\end{thebibliography}
